\documentclass{emulateapj}
\usepackage{apjfonts}
\usepackage{lscape}
\newcommand{\rotate}{}

\shortauthors{Becker et al.}
\shorttitle{Variable Galactic Radio Sources}

\slugcomment{Accepted for publication in the Astronomical Journal}

\begin{document}
\title{Variable Radio Sources in the Galactic Plane}

\author{
Robert~H.~Becker\altaffilmark{1,2},
David J. Helfand\altaffilmark{3,4},
Richard~L.~White\altaffilmark{5},
Deanne~D.~Proctor\altaffilmark{2}
}
\email{bob@igpp.ucllnl.org}

\altaffiltext{1}{Physics Dept., University of California, Davis, CA 95616}
\altaffiltext{2}{IGPP/Lawrence Livermore National Laboratory}
\altaffiltext{3}{Dept. of Astronomy, Columbia University, New York, NY 10027}
\altaffiltext{4}{Quest University, Squamish, BC V8B 0N8}
\altaffiltext{5}{Space Telescope Science Institute, Baltimore, MD 21218}

\begin{abstract}

Using three epochs of VLA observations of the Galactic Plane in the
first quadrant taken $\sim15$ years apart, we have conducted a
search for a population of variable Galactic radio emitters in the
flux density range 1--100 mJy at 6~cm. We find 39 variable sources
in a total survey area of 23.2 deg$^2$. Correcting for various
selection effects and for the extragalactic variable population of
active galactic nuclei, we conclude there are $\sim1.6$ Galactic
sources deg$^{-2}$ which vary by more than 50\% on a time scale of
years (or shorter). We show that these sources are much more highly
variable than extragalactic objects; more than 50\% show variability
by a factor $>2$ compared to $<10\%$ for extragalactic objects in
the same flux density range. We also show that the fraction of
variable sources increases toward the Galactic center (another
indication that this is a Galactic population), and that  the
spectral indices of many of these sources are flat or inverted. A
small number of the variables are coincident with mid-IR sources
and two are coincident with X-ray emitters, but most have no known
counterparts at other wavelengths. Intriguingly, one lies at the
center of a supernova remnant, while another appears to be a very
compact planetary nebula; several are likely to represent activity
associated with star formation regions. We discuss the possible
source classes which could contribute to the variable cohort and
followup observations which could clarify the nature of these
sources.

\end{abstract}


\section{Introduction}

Variable radio emission is a hallmark of energetic objects such as
coronally active stars, supernovae, neutron stars, black holes, and
active galactic nuclei (AGN). Indeed, radio variability is often
indicative of high-energy processes and, in principle, can be
valuable for finding examples of relatively rare objects. However,
surveys for variability are themselves quite rare --- blind sky
surveys are almost never repeated owing to the scarcity of telescope
time. The exceptions are mostly in the optical regime. Comparisons
between POSS1, POSS2, and SDSS have been useful for studying
variability (de~Vries et al.\ 2005). Gravitational microlensing
studies (e.g., Alcock et al.\ 1997) and supernova searches (e.g.,
Astier et al.\ 2006, Miknaitis et al.\ 2007) have produced a wealth
of data on optical variability from targeted sky regions, and several
upcoming experiments such as Pan-STARRS (Kaiser et al.\ 2002), the
Palomar Transit Factory (Rau et al.\ 2009), and LSST (Tyson 2002)
will make the coming decade one in which time-domain astronomy plays
a prominent role.

Variability studies in the radio band have typically targeted bright
extragalactic sources (see de~Vries et al.\ 2004 for a review of searches
for, and mechanisms of, radio variability). Comparisons between blind
radio surveys are often hampered by differences in angular resolution and
the confusing presence of interferometric sidelobe patterns. For example,
there has been no systematic search for radio variability between the
two largest radio sky surveys, FIRST (Becker et al.\ 1995) and NVSS (Condon
et al.\ 1998). The FIRST survey did observe one area twice at 1400 MHz, an
equatorial strip $\sim1.5$ degrees wide in the range
$21^{\mathrm h}20^{\mathrm m} < RA < 03^{\mathrm h} 20^{\mathrm
m}$. A search for variable sources in this area was reported in
de~Vries et al.\ (2004). The search covered $\sim120$~deg$^2$ of
extragalactic sky with a sensitivity similar to the Galactic plane
search reported here; it thus serves as a useful control from which
to estimate how many of the sources we find are background extragalactic
radio sources.

The most systematic search for radio variablity in the Galactic plane
used the NRAO 91-m telescope in Greenbank, WV, operating at a
frequency of 5~GHz (Gregory \& Taylor 1986). Over a five-year period
the plane was observed 16 times, leading to the detection of 32
variable radio sources. The survey had a flux density threshold of
$\sim20$~mJy and an angular resolution of 3\arcmin. Using the
Very Large Array\footnote{The
Very Large Array is an instrument of the National Radio Astronomy
Observatory, a facility of the National Science Foundation operated
under cooperative agreement by Associated Universities, Inc.}
(VLA), the Galactic plane has been surveyed at this same frequency
but with much higher sensitivity and angular resolution, although
with minimal repetition (Becker et al.\ 1994).

Recently, a new Galactic plane survey at 6~cm (4.86~GHz) has begun
at the VLA. The new survey (CORNISH\footnote{The Co-Ordinated Radio
'N' Infrared Survey for High-mass star formation; see
\url{http://www.ast.leeds.ac.uk/Cornish}}; Purcell et al.\ 2008)
has substantial overlap with our previous survey; both surveys have
a flux density threshold of $\sim1$~mJy. Comparison between these
two data sets allows for a search for Galactic radio sources that
exhibit variability over the fifteen-year interval between the
surveys. The search is complicated because the two surveys use
different VLA configurations and hence have different angular
resolutions (5\arcsec\ versus 1.5\arcsec).
Nonetheless, it is possible to identify strongly varying sources.
In this paper we will compare results between the original survey
and two epochs of data from the new survey. In section 2 we discuss
the parameters of the two surveys, while in section 3 we present
the results from a comparison of the two samples and adduce evidence
for variability. We describe the properties of the variable sources including
their spatial distribution, spectral indices, and counterparts at other wavelengths
(\S4) and end with a discussion of our limited knowledge of the nature of
these objects (\S5).

\begin{figure*}
\epsscale{1.0}
\plotone{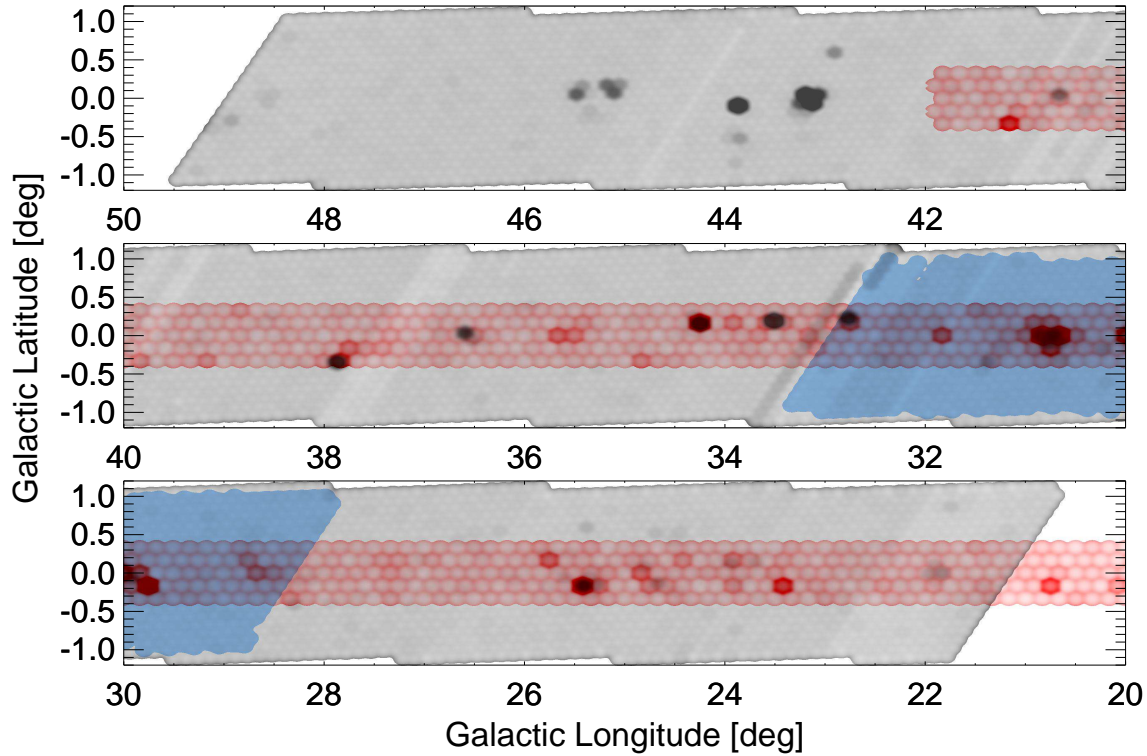}
\caption{
Sky coverage for overlapping regions from the 1990${+}$ (red), 2005
(blue), and 2006 (gray) 6~cm survey epochs. Darker regions have
higher rms noise values, while white areas are outside the survey.
The noise is higher at the edges of the coverage and in fields with
bright or complex extended sources.  Typical rms values are $\sim
0.1$~mJy in both surveys, but the old 6~cm data were acquired with
a more widely spaced pointing grid and so display greater variation
with position. The areas of overlap are given in Table~1.
}
\label{fig-coverage}
\end{figure*}

\section{The 6~cm Surveys}

The original VLA 6~cm Galactic plane survey was carried out between
1989 and 1991 in the C and BnC configurations (Becker et al.\ 1994).
It covered a longitude range $-10^{\circ} < l < 42^{\circ}$ within
$\pm 0.4^{\circ}$ of the plane for a total of 43~deg$^2$. The data
were re-reduced in 2005 using much improved data processing algorithms
and some additional data (White et al.\ 2005).  The new catalog
reaches a flux density limit of $\sim1$~mJy and contains over 2700
radio sources. Since the data were taken in C and BnC configurations,
the angular resolution is $\sim5\arcsec$.

The new CORNISH survey (Purcell et al.\ 2008) is meant to complement
the {\it Spitzer} GLIMPSE Legacy program (Benjamin et al.\ 2003).
When completed, it will cover a longitude range $10^{\circ} < l <
65^{\circ}$ within $\pm 1^{\circ}$ of the plane. The data are being
taken in the B configuration and hence will have an angular resolution
of $\sim1.5\arcsec$. The new survey will also achieve a flux
density sensitivity of $\sim1$~mJy. The ultimate areal coverage
will be 110~deg$^2$. A pilot study of 10~deg$^2$ near $l = 30^{\circ}$
was carried out in the spring of 2005. The first 64~deg$^2$ of the
full survey (including repeated observations of the pilot area) was
observed in summer 2006.
We retrieved these data from the VLA archive and reduced them using
the AIPS procedures we developed for the FIRST survey (White et
al.\ 1997). Our source detection algorithm HAPPY was run on the
final co-added images.

We henceforth refer to the three epochs as~I (1990${+}$), II~(2005)
and III~(2006).  Note that while the epoch~II and~III data were
taken over short periods of time (spanning about 2 months in each
case), the epoch~I data were taken over a much greater time period (hence
our choice of the label ``1990${+}$'').
For the overlapping area used for this paper, 70\% of the epoch~I
observations were taken between June 1989 and December 1990, and
30\% were taken between February and April of 2004.  Consequently
the time span between epoch~I and the later two epochs varies by a
large factor depending on the source location.  We will report the
mean observational epoch of the flux density measurements for
individual objects in the following discussion.

For this paper we restrict our attention
to sources in sky regions with coverage at two or three epochs.
Table~1 describes the areas of overlap between the various epochs
and the number of sources from each catalog included in those areas.
To ensure source reliability, we restrict our sample to sources
that are strongly detected ($>8.5\sigma$) in one of the epochs or
that are confirmed by detections at multiple epochs.  We also check
for detections at 20~cm, either from our MAGPIS survey (Helfand et
al.\ 2006) or in the catalog of White et al.\ (2005); a 20~cm detection is
required as confirmation for sources detected in only one 6~cm epoch.

In
Figure~\ref{fig-coverage} we show the sky coverage for the three
epochs in the vicinity of the overlap region.  Note that the 2005
pilot area is entirely covered by the 2006 data, so all of the sky
area covered by 1990${+}$ and 2005 observations also has 2006
observations.

\begin{deluxetable}{ccccc}
\label{table-area}
\tablecolumns{5}
\tablewidth{0pc}
\tabletypesize{\scriptsize}
\tablecaption{Sky Regions with Multiple Epochs of Observations}
\tablehead{
\colhead{Epochs} & \colhead{Area Covered} & \multicolumn{3}{c}{Number of Sources} \\
& \colhead{(deg$^2$)} & 1990${+}$ & 2005 & 2006 \\
\colhead{(1)} & \colhead{(2)} & \colhead{(3)} & \colhead{(4)} & \colhead{(5)}
}
\startdata
1990${+}$, 2006 & 13.4 & 541 & \nodata & 347 \\
2005, 2006 & \phn5.9 & \nodata & 144 & 142 \\
1990${+}$, 2005, 2006 & \phn3.9 & 161 & 168 & 133 \\
\enddata
\end{deluxetable}

The other significant difference between the two surveys is their
angular resolution. For unresolved radio sources, the flux densities
from the two surveys are directly comparable; the difficulty comes
in knowing which sources are true point sources. For sources that
are partially resolved by the new survey, the flux density will be
lower than that measured a decade and a half ago, even in the absence
of variability. Hence, partially resolved sources will give a
false-positive variability signal. By the same token, however, any
source significantly brighter in the newer survey is almost certainly
variable. All images from both surveys can be found at the MAGPIS
website (\url{http://third.ucllnl.org/gps}).

\section{Search for Variability}

A match among the three 6~cm data sets resulted in 503 distinct
sources detected at two or more epochs.
To ensure reliability,
we restrict our sample to sources that are detected in at least two 6~cm
epochs or that have confirming detections at 20~cm.  Sources detected only
in a single 6~cm epoch and not at 20~cm are excluded.
Sources are considered a
match if their positions agree to within 1.5\arcsec\ for the epoch~II
and~III catalogs or to within 5\arcsec\ between the epoch~I and
later-epoch catalogs.  These relatively large match radii are chosen
to include extended sources, which can have larger positional
offsets.  For the higher resolution epoch~II and~III data, the
median position difference is 0.2\arcsec, and 80\% of the sources
have positions that agree to within 0.4\arcsec.  For comparisons
between the low-resolution epoch~I data and the more recent
observations, the median separation is 0.7\arcsec, and 80\% of the
sources have positions that differ by 1.6\arcsec\ or less. To avoid
potential confusion, we have removed from the match list sources
that have ambiguous matches owing to multiple components within the
matching radius.

\begin{figure*}
\epsscale{0.8}
\plotone{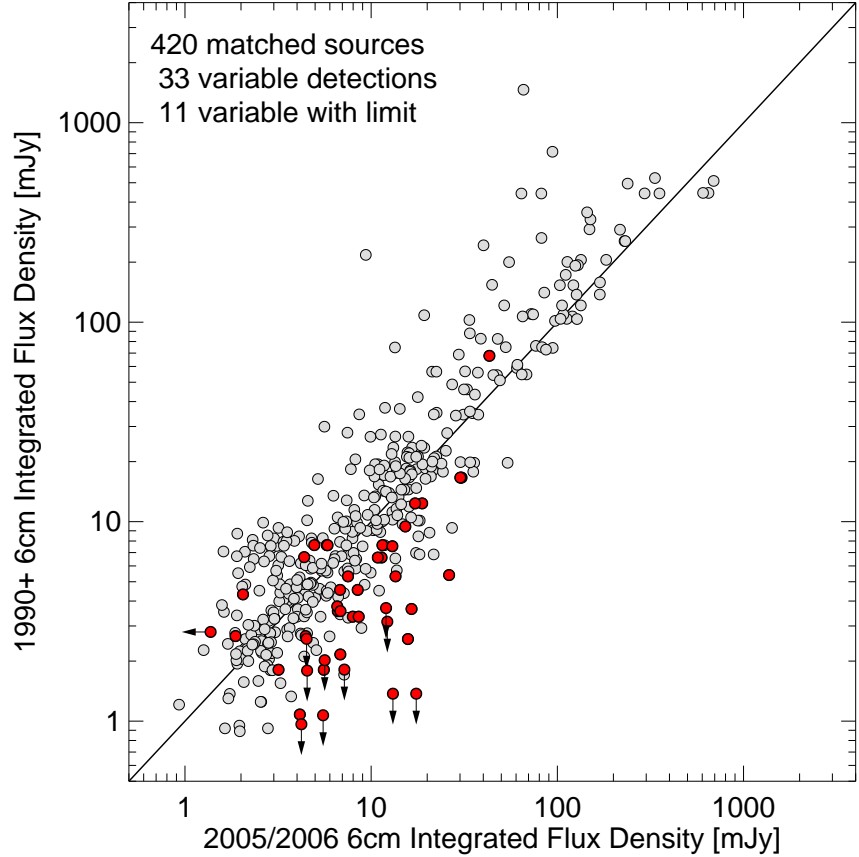}
\caption{
Comparison of 6~cm integrated flux densities for old (epoch I/1990${+}$) and
new (epochs II/2005 and III/2006) catalogs. Source in all three epochs are plotted
twice to show both the 2005 and 2006 flux densities.
Red symbols indicate the variable
sources, with upper limits shown for variables detected at only one
epoch. Most sources have similar flux densities in the two epochs,
but extended sources tend to have lower flux densities in the newer
survey because those data were taken in a higher-resolution VLA
configuration that resolves out some of the extended radio emission.
Consequently, in our variable source search there is a bias toward
objects that are brighter in the 2005/2006 epoch.
}
\label{fig-flux}
\end{figure*}

A comparison of the flux densities determined from the old and new
data is plotted in Figure~\ref{fig-flux}; sources that fall along
the diagonal have comparable flux densities from the two measurements.
There is a clear bias for sources to be weaker in the newer
observations, a direct consequence of the higher angular resolution,
which results in slightly extended sources having some of their
flux resolved out in the newer data.  Certainly some of these sources
could be variable, but it is difficult to distinguish between a
decrease due to variability and a decrease due to resolution effects.
Happily, the reverse is not true; sources that brighten between the
two epochs are likely to be truly variable.

\begin{figure*}
\epsscale{0.8}
\plotone{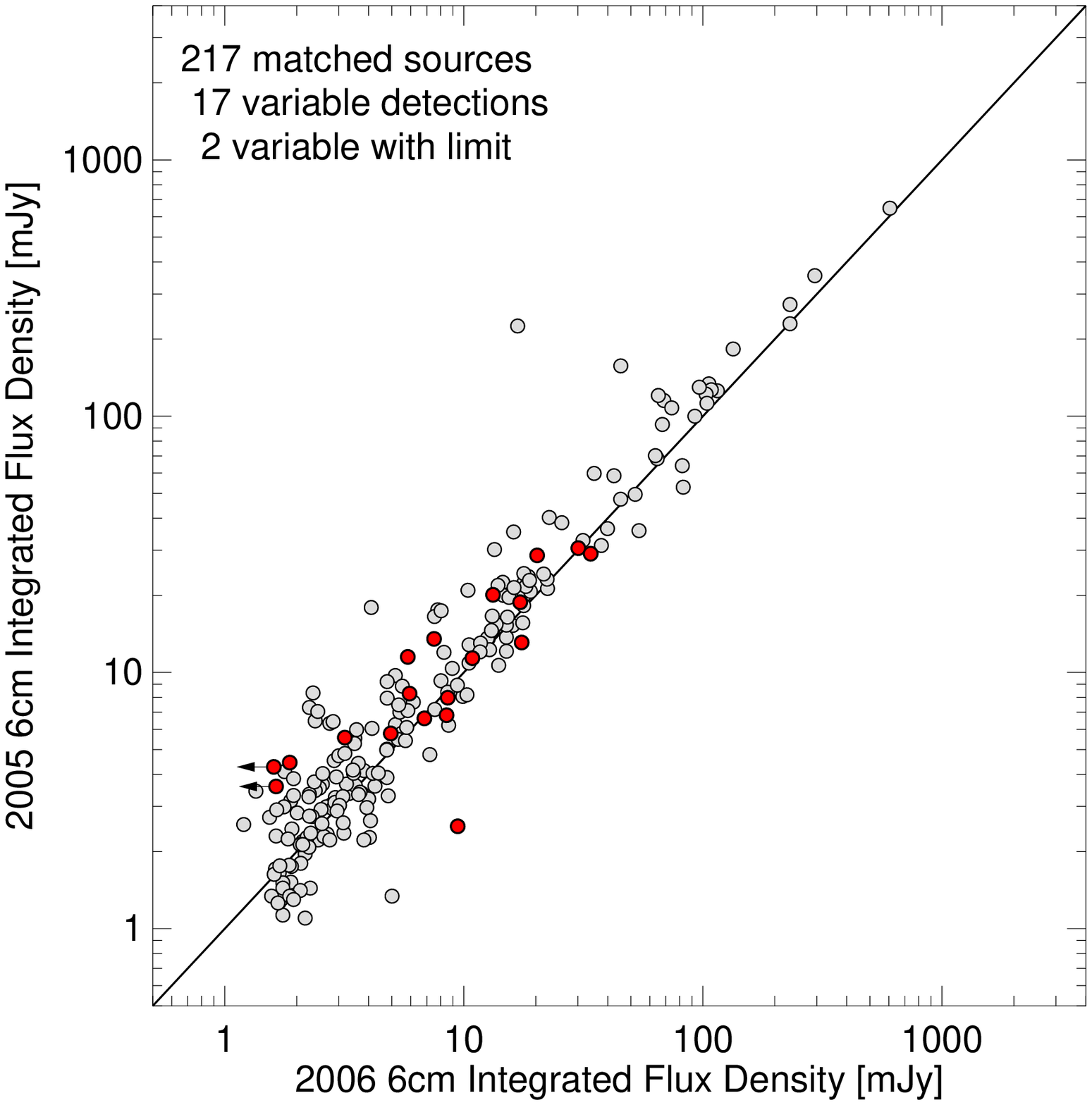}
\caption{
Comparison of the integrated flux densities for the epoch~II and~III
catalogs.  These observations were taken in the same VLA configuration,
which makes the flux densities directly comparable. The two epochs
are also much closer in time, reducing the amplitude of the expected
variability signal.  The symbols are the same as in Fig.~2.
}
\label{fig-flux2}
\end{figure*}

Figure~\ref{fig-flux2} displays a comparison of the flux densities
measured in the two high-resolution epochs (II and~III).  The area
of overlap is smaller (5.9~deg$^2$ versus 17.3~deg$^2$ for
Fig.~\ref{fig-flux}), but it is clear that the scatter is considerably
reduced.  This is expected because the observations are taken in
the same VLA configuration and so have the same resolution.  (The
shorter time baseline for variability also presumably contributes
slightly to the reduced scatter.)

The sources falling in a region covered by at least two of the three
catalogs yielded a list of potential variable sources using a
$5\sigma$ variability threshold\footnote{The difference in the peak
flux densities was required to be greater than
$5 \times (\sigma_{\rm old}^2 + \sigma_{\rm new}^2)^{1/2}$ if the source
was detected at both epochs; if the object was detected at only one
epoch, the flux density at the undetected epoch was conservatively
taken to be twice the rms value at that epoch.}.  A visual inspection
of the pairs of images led us to reject many as suspect owing either
to source confusion or to clear angular extent in the higher
resolution observations.  There remained 39 sources regarded as
having a high likelihood of being true radio variables (Table~2).
We were cautious about including sources that were brigher in epoch~I
due to the resolution difference discussed above.  Only 5 of the
candidates rely on a bright epoch~I measurement to establish
variability; the other 34 either are brighter in the high-resolution
data or show variability between epochs~II and~III.  We have retained
the five sources that fade from their epoch~I flux measurements
because they appear to be point-like in all epochs at both 6~cm and 20~cm,
but we cannot
categorically exclude a small source extent being responsible for
the lower flux density observed in the more recent data.  The
variability significance in column 16 of Table~2 is shown in bold
italic type for these less reliable sources.

In examining candidate variables, we were alert to the possibility
of calibration errors or bad data causing systematic flux density
differences.  Consequently we looked carefully at the close pair
of sources, G37.7347${-}$0.1126 and G37.7596${-}$0.1001, both of
which were undetected in epoch~I and were bright ($>10$~mJy) in
epoch~III.  To confirm the reality of these sources, we examined
the individual grid images that contributed to the coadded images
for each source.  The grid images confirmed the variation: in epoch~I
each source would have been detected at more than $5\sigma$
significance in two different grid images if the source was as
bright as in epoch~III, but neither showed any evidence for emission.
And in epoch~III the sources were detected in two or more independent
observations at flux densities consistent with the coadded image
detection.

Of the five single-epoch detections in the list, one appears only
in epoch~I (and so is one of the less reliable sources), one appears
only in epoch~II, and three appear only in epoch~III.  Note that
single-epoch sources both must be relatively bright in the detected
epoch for the variability to be considered and also must have
confirming detections at 20~cm.  In fact, all of the multi-epoch
sources are also detected at 20~cm; we use the spectral indices
derived from the 6 and 20~cm flux densities below, although caution
is warranted since the 20~cm and 6~cm observations are not
contemporaneous, so the variability for which we are selecting will
also affect the spectral index estimates.  In fact, roughly half
of the MAGPIS 20~cm flux densities are inconsistent with our original
compact source survey at this wavelength undertaken in the 1980s
(see catalogs in White et al.\ 2005), underscoring the case for
variability. The degree of variability at 6~cm ranges from 20\% to
a factor of 18. The flux density distribution ranges from $<1$ to
65~mJy with a median of $\sim8$~mJy in the second-epoch data.

\section{Characteristics of the Variable Sources}

Only a few papers have reported variability results for
centimetric radio sources from extragalactic surveys
on time scales of years and with sensitivities
in the mJy range. Bower et al.\ (2007) examined archival VLA data
from a frequently observed calibration field (944 observations over
22 years) to search for transient radio sources at 6~cm to a flux
density limit of 370~$\mu$Jy; a transient source was defined as one
that only appeared at a single epoch (or over a short range of
contiguous epochs $<2$ months in length). They measured a two-epoch rate of
$1.5\pm0.4$ transients per square degree, and
estimated that the number of transients scales with flux density
as $S^{-1.5}$. Our faintest variable source has a 6~cm flux density
of 2.8~mJy (at the brighter epoch; fainter objects would not have
passed the $5\sigma$ variability threshold). Thus, we should expect
$\sim1$ true extragalactic transient in our survey area of
23.2~deg$^2$ (the total pair-wise area covered -- see Table~1). While
there are five sources detected at one epoch only in our sample,
all four are detected independently at 20~cm (at one or more different
epochs), and thus cannot be considered true transients. The detection
of zero transients when one is expected is construed as consistent
with the results of Bower et al.\ (2007) while setting a weak limit
on the number of Galactic transient sources.

Examining data collected for a deep survey of the Lockman Hole
spaced at intervals of 19 days and 17 months, Carilli et al.\ (2003)
concluded that only 2\% of sources between 50 and 100~$\mu$Jy at
1.4~GHz are highly variable ($>50$\%). These observations did,
however, yield nine variable objects in the flux density range 1
to 25~mJy in a field with a FWHM 32\arcmin; only one of these
sources varied by more than 50\%. 

The survey by de~Vries et al.\ (2004) mentioned in the introduction
(\S1) offers the best comparison sample against which to assess the
fraction of our variable sources likely to be extragalactic. They
find 123 sources variable at $>4\sigma$ significance in 120.2~deg$^2$
of high-latitude sky, or roughly 1 variable source per square degree.
The median flux density of the extragalactic sample is 13.5~mJy at
$\lambda=20$~cm. As noted above, we have 20~cm flux densities (albeit
non-contemporaneous ones) for all members of our sample; the median
flux density is 12.9~mJy, very similar to that of the de~Vries
sample.

\begin{deluxetable}{ccccc}
\label{table-vardist}
\tablenum{3}
\tablecolumns{5}
\tablewidth{0pc}
\tabletypesize{\scriptsize}
\tablecaption{Distribution of Variability}
\tablehead{
\colhead{Fractional} & \colhead{de~Vries} & \colhead{This Paper} & \colhead{Predicted} & \colhead{Galactic}\\
\colhead{Variability $f$} & \colhead{Counts} & \colhead{Counts} & \colhead{Extragalactic} & \colhead{Excess} \\
\colhead{(1)} & \colhead{(2)} & \colhead{(3)} & \colhead{(4)} & \colhead{(5)}
}
\startdata
  $<$ 1.25   &    51 &      1 & 7.4 &   \nodata \\
1.25 -- 1.50 &    39 &      5 & 5.7 &   0.0 \\
1.50 -- 1.75 &    19 &      5 & 2.8 &   2.2 \\
1.75 -- 2.00 & \phn4 &      4 & 0.6 &   3.4 \\
2.00 -- 2.25 & \phn4 &      2 & 0.6 &   1.4 \\
2.25 -- 2.50 & \phn2 &      3 & 0.3 &   2.7 \\
2.50 -- 2.75 & \phn1 &      2 & 0.1 &   1.9 \\
2.75 -- 3.00 & \phn1 &      0 & 0.1 &   0.0 \\
  $>$ 3.0    & \phn2 & 17\phn & 0.3 &  16.7\phn \\[\smallskipamount]
Total        & 123\phn & 39\phn & 17.9\phn &  28
\enddata
\tablecomments{
Col.~(1): Ratio of brightest to faintest flux density measurements.
Col.~(2): Variable source counts from de~Vries et al.\ (2004).
Col.~(3): Variable source counts from this paper.
Col.~(4): de~Vries counts scaled to match area covered in this paper.
Col.~(5): Net excess of variables in this paper compared with de~Vries counts.
}
\end{deluxetable}

However, an examination of the fractional variability of the two
samples (defined as $f$, the highest flux density recorded over the
lowest) reveals drastic differences. Table~3 displays the distribution
of fractional flux density variation for the de~Vries
extragalactic sample and for our Galactic plane catalog of
variables. A total of $73\pm4$\% of the extragalactic sample has a
fractional variability of $f<1.5$ while only 6 of the 39 Galactic
variable (15\%) vary this little. At the other end of the distribution,
only 2/123 extragalactic objects vary by as much as a factor of 3,
while fully 17/39 (44\%) of the Galactic plane sources are this
variable.

We can use the de~Vries sample to estimate the number of extragalactic
variables present in our survey area. The ratio of areas is 23.2
deg$^2$/120.2 deg$^2$ or 0.193. We cannot simply scale by area,
however, because of the resolution bias, discussed above, that
discriminates against sources which faded between epoch~I and
epochs~II and~III. Of the 30 sources whose variability was established
on the basis of a change between epoch~I and a later epoch, five
sources faded and 25 sources brightened in the later epochs. Since
the distribution should be inherently symmetrical, we can assume
we eliminated roughly 20 fading sources to protect against resolution
effects. Thus, the total number of true variables is reduced by
20/50 or 40\%; note that this correction factor applies only to the
area in which 1990${+}$ data is compared to later data (19.3 deg$^2$).
The effective sky area covered by our survey when this inefficiency
is taken into account is $A_{\rm eff} = 0.6 \times 19.3 + 5.9 =
17.5\,\hbox{deg}^2$, and the expected number of extragalactic
variables is therefore $123 \times A_{\rm eff}/120.2 = 18$.

If we distribute these 18 sources with the fractional variability
of the extragalactic sample (column 4 of Table~3), we expect
$\sim13$ at $f<1.5$, 3 with $1.5<f<2.0$, and $\sim2$ to vary by
more than a factor of 2.  The distribution for Galactic variables
is 6 at $f<1.5$, 9 with $1.5<f<2.0$, and 24 varying by more than a
factor of two.  Our Galactic sample includes few sources that vary
by less than $25\%$ because the complexity of radio emission in the
Galactic plane combined with differences in the VLA configuration
required a higher threshold for confident detection of variability;
that is the reason for our use of a $5\sigma$ threshold rather than
the $4\sigma$ threshold adopted by de~Vries et al.  Ignoring this
lowest bin as below our survey sensitivity, the de~Vries catalog
has 72 variable sources, yielding an expectation of $\sim10.5$
sources in our area.  Subtracting these numbers from our catalog
suggests that while many of the sources at $f<1.5$ are extragalactic,
only $\sim5$ of the 33 sources with greater variability are background
objects.  Scaling by the effective area $A_{\rm eff}$ derived above,
we find a surface density of Galactic variable sources of 1.6~deg$^{-2}$,
nearly six times the 0.3~deg$^{-2}$ density of extragalactic
sources that vary by more $50\%$ on time scales of years.

A complicating factor in this comparison is that the de~Vries
variables were selected at 20~cm rather than 6~cm wavelength.  The
most variable AGN have beamed nuclear radio emission that usually
has a flatter spectral index than the non-variable extended radio
emission. Consequently at a shorter wavelength, the beamed emission
will be more dominant, which will increase the apparent amplitude
of variability.  Without a good extragalactic comparison sample at
6~cm, it is difficult to determine exactly how large this effect
will be.  But we consider it very unlikely that the changes resulting
from the wavelength difference could explain the large differences
seen between the extragalactic and Galactic samples in the frequency
of large-amplitude variables.

\begin{figure*}
\epsscale{0.9}
\plotone{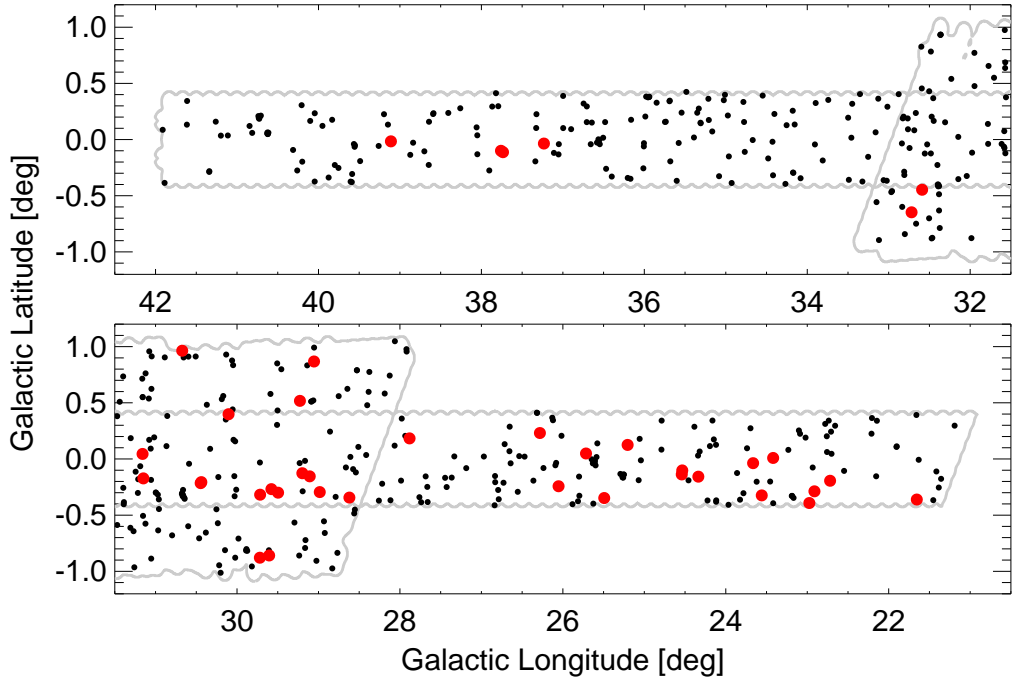}
\caption{
Distribution of variable sources (red) compared with
the population of sources that are point-like at
6~cm and that either are detected at multiple
epochs or are detected at greater than
$8.5\sigma$ significance in a single epoch.
}
\label{fig-posdist}
\end{figure*}

\begin{figure*}
\epsscale{0.6}
\plotone{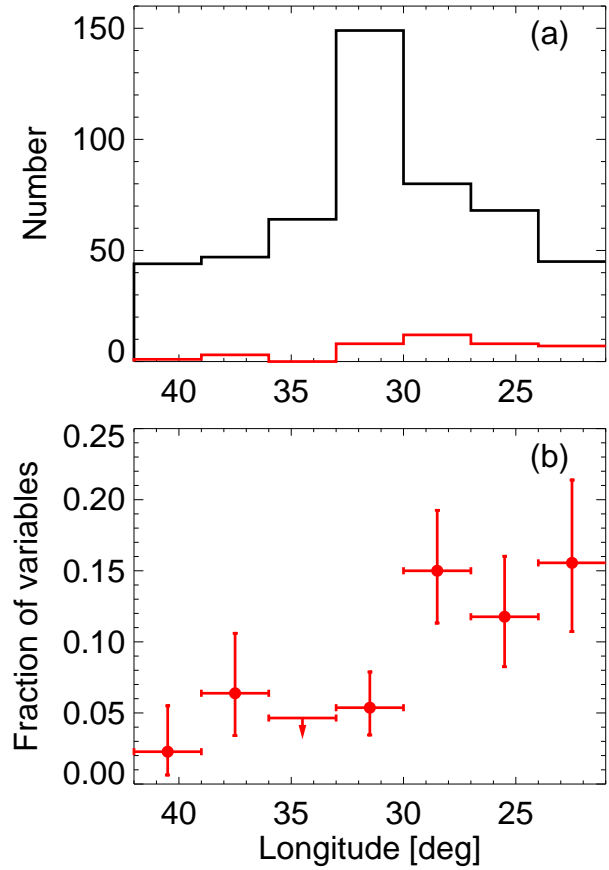}
\caption{
Longitude distribution of variable sources (red) compared with
the same comparison sample as in Fig.~4.  The top panel (a) shows
the raw counts, while the bottom panel (b) shows the fraction of
variable sources in each bin, with error bars computed from the binomial
probability distribution (and a $2\sigma$ upper limit).  The variability
fraction increases dramatically toward the Galactic center, which
indicates that the variable sources are mainly a Galactic
rather than extragalactic population.
}
\label{fig-longhist}
\end{figure*}

The spatial distribution of the sources in our catalog strongly
supports the view that they are dominated by a Galactic population.
In Figure~\ref{fig-posdist} we compare the distribution of variable
sources with a parent population consisting of point-like sources
(defined by an integrated flux density less than 25\% larger than
the peak flux density) that either are detected at multiple 6~cm
epochs or that are strong single-epoch detections ($>8.5\sigma$).
Such sources could have been detected as variables.  The latitude
distribution shows a bias toward negative latitudes, consistent
with earlier studies that show that $b=0.0$ lies above the Galactic
plane in the first quadrant. 
In Figure~\ref{fig-longhist}, we display the
longitude distribution for all sources (which is distorted by
differential coverage) and for the variable sources. The fraction
of variables, displayed in the lower panel, shows a clear rise
toward the Galactic center.

\begin{figure*}
\epsscale{0.6}
\plotone{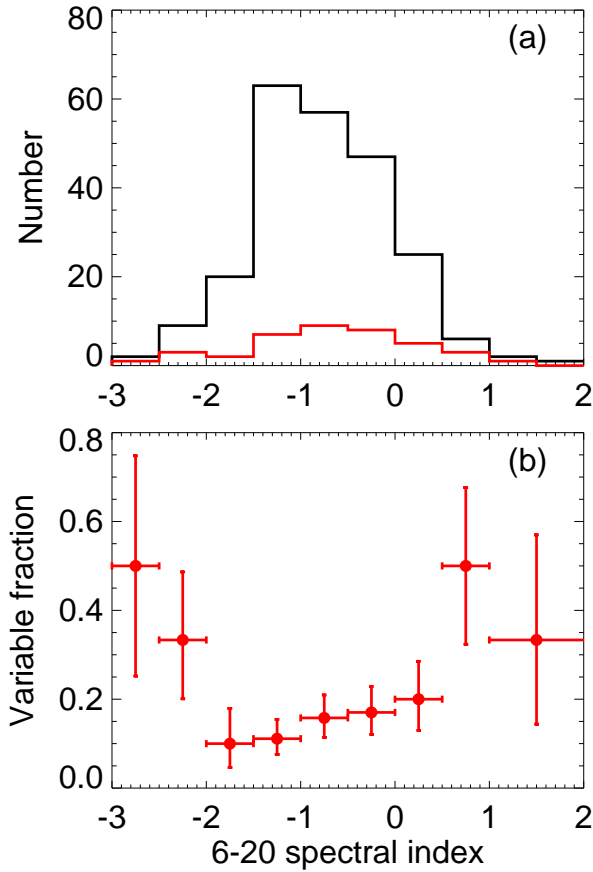}
\caption{
Variability fraction as a function of the spectral index $\alpha$ (
$F_\nu \sim \nu^\alpha$) between 6~cm and 20~cm.
The comparison sample includes the sources from
Figs.~4 and 5 that have 20~cm flux measurements.  The spectral
index was computed from the lowest 6~cm flux density at any epoch
in order to reduce the selection bias toward flatter spectra in variable
sources.
}
\label{fig-spindhist}
\end{figure*}

The spectral index distribution (Figure~\ref{fig-spindhist}) also
suggests that the variable sources represent a distinct population.
There is a trend toward greater variability as the radio spectrum
becomes flatter (increasing spectral index).  For any individual
source, the spectral index calculated between our 6~cm and 20~cm
catalogs is unreliable, as the measurements were 1) obtained with
different spatial resolutions, and 2) far from contemporaneous.
Since the 20~cm observations have lower resolution, they detect
more flux in extended sources and so tend to produce spectral indices
that are too steep.  On the other hand, the observed variability
in the 6~cm flux density tends to bias the index toward flatter
values, since sources that brighten at 6~cm are more likely to be
recognized as variable.  To compensate partially for the latter
effect, the spectral index in Figure~\ref{fig-spindhist} is computed
using the smallest 6~cm flux measured at any of the three epochs.  That
may also be responsible for the variable sources with very steep
spectral indices ($\alpha < -2$), which may have been in a bright
phase when measured at 20~cm.

The interpretation of the spectral index distribution is therefore
not straightforward.  The apparent increase in variability for
flatter spectrum sources could result from either a Galactic
population (optically thin or thick thermal emission) or 
an extragalactic population (beamed emission from AGN/blazars).

Another line of evidence that many of the variable sources are
Galactic derives from their counterparts at other wavelengths.  We
examined images from the {\it Spitzer} GLIMPSE survey (3.6, 4.5,
5.8 and 8.0$\mu$m; Benjamin et al.\ 2003), the {\it Spitzer} MIPSGAL
24$\mu$m survey (Carey et al.\ 2009), and the 1.1~mm Bolocam Galactic
Plane Survey (BGPS; Aguirre et al.\ 2009).  Of the 39 variable
sources, 7 are MIPSGAL sources, with 6 of those also found to be
GLIMPSE sources, and 4 are detected in the BGPS millimeter observations;
all are described in greater detail below.  None of the counterparts
are expected to be the result of chance coincidences, implying that
all of these objects must be in the Galaxy.  Infrared/mm counterparts
are significantly more common among the variable sources than
among the non-variable 6~cm sources.  We examined a sample of 40
non-variable radio sources, selected as unresolved sources detected
in at least 2 epochs with 6~cm flux densities that are consistent
within $2\sigma$ at all epochs.  Only 2 of the non-variable objects
were found to have MIPSGAL counterparts, and none had BGPS matches.
We conclude that the existence of these counterparts is related to
the nature of the variable sources.

\section{Discussion}

\subsection{Source Identification -- What They Are Not}

Having established the existence of a population of highly variable
Galactic sources, the obvious question is, What are they? Three
classes of Galactic variable radio emitters can be easily eliminated from
consideration: coronally active radio-emitting stars, pulsars, and
masers. We justify our exclusion of these source classes in turn.

In a survey of 122 RS CVn and related active binary systems which
are among the most luminous stellar radio sources, Drake et al.
(1989) found only 18 detected above a quiescent flux density of
1~mJy at 6~cm; the faintest optical counterpart was $V=10.0$. Even
assuming an extreme flare of a factor of 100 (Osten 2008), the
faintest possible counterpart would have $V=15$; none of our variables
has a counterpart this bright. As for dMe flare stars, the other
main class of variable stellar radio emitters, the most luminous
quiescent emission is $\sim10^{14.2}$ erg s$^{-1}$ Hz$^{-1}$ (Gudel
et al.\ 1993) corresponding to a flux density of $\sim1$~mJy at a
distance of 13~pc. Even an extreme flare with an increase of a
factor of 500 over the quiescent level (Osten 2008 and references
therein) would fall below our flux density threshold for a distance
$>290$~pc. Stars with spectral types later than M6 would have
counterparts fainter than 20th magnitude and could be represented
in our sample. However, statistically, M-stars cannot be a significant
contributor; Helfand et al.\ (1999) found only $\sim5$ M stars in
5000 deg$^2$ of the {\sl FIRST} survey to a flux density limit of
0.7~mJy, whereas our variables have a surface density of 1.6~deg$^{-2}$.

While nearby radio pulsars scintillate strongly in the ISM leading
to large-amplitude variability, pulsars have very steep radio
spectra, and most have not been detected at 6~cm (none of our objects
are coincident with one of the 1827 known pulsars; Manchester et
al. 2005). For a typical spectral index of $-1.5$, our weakest
source would be a $\sim100$~mJy pulsar at 400~MHz, and most unlikely
to have been missed in pulsar surveys.  The small duty cycle of the
recently discovered RRATs (Rotating RAdio Transients -- McLaughlin
et al.\ 2006) makes them equally unlikely to explain our variable
sources.

Finally, radio masers are known to be highly variable, but no known
maser transitions fall within our bandpass. As noted below, however,
three of our variables are coincident with methanol masers.

Two classes of extragalactic radio transients --- supernovae and GRB
afterglows --- are also highly improbable counterparts for our events.
Both have rise times of at most tens of days and cannot, in the
absence of a steady underlying source of radio emission, account for
the bulk of our sources which show a flux density increase over
many years. In addition, their rarity makes them statistically
unlikely counterparts. The one extragalactic population that does
show variability on the time scales we probe, AGN, are shown above
to have variability amplitudes which exclude them from explaining
all but a handful of our events.

The remaining known classes of variable radio sources include
microquasars (accreting, high-mass X-ray binaries that produce
relativistic jets: e.g., SS433, Cyg X-3 and GRS 1915${+}$105), radio
magnetars (Camilo et al.\ 2006; Camilo et al.\ 2007), and the recently
described Galactic Center Transient sources (Hyman et al.\ 2009 and
references therein).  The first two of these have signatures at
other wavelengths; we explore below the fragmentary data outside
the radio band that is available for our variable objects.

\subsection{Source Identification -- Multiwavelength Data}

Counterparts at other wavelengths can be useful in suggesting the
origin of radio variability. At our MAGPIS website (Helfand et al.
2006), we have collected the following Galactic plane data in
addition to the three-epoch 6~cm data described herein: two epochs
of 20~cm observations for these same fields including the principal
MAGPIS survey, 90~cm observations of the same regions, the 3.6,
4.5, 5.8, and 8.0 $\mu$m data from the Spitzer {\sl GLIMPSE} survey
(Benjamin et al.\ 2003), 24 $\mu$m images from MIPSGAL (Carey et
al.\ 2009), 20 $\mu$m data from the MSX survey (Price et al.\ 2001)
and the 1.1~mm Bolocam Galactic plane survey (Aguirre et al.\ 2009).
In addition, we have queried the SIMBAD database for each of our
sources and have examined the Digitized Sky Survey images; in one
case, we have obtained optical observations of a source. We report
the results of this multi-wavelength inquiry here.

\subsubsection{Mid-IR and mm observations}

Seven of our variables are detected at 24 $\mu$m in the MIPSGAL
survey, and six of these are also detected in at least one GLIMPSE
mid-IR band. Four of the objects are also detected at 1.1~mm in the
Bolocam survey. In all seven cases at least two bands are available,
and in all seven cases the sources are red; i.e., they are faintest
in the short-wavelength bands and brightest in the long-wavelength
bands. In two cases (G31.1595${+}$0.0449 and G37.7347${-}$0.1126) multiple
components with different IR spectral shapes are present, with the
radio source identified with the brighter component in the first
case, and the redder component in the second. Three of the IR-detected
objects have associated methanol masers; this, coupled with their
IR spectra demonstrate they represent activity associated with
star formation in compact or ultracompact \ion{H}{2} regions.

For one IR-detected source, G29.5779${-}$0.2685, we have obtained
followup observations at the MDM Observatory (J.~Halpern, private
communication).  R-band and H$\alpha$ images were obtained on
23~August 2009, and show a barely resolved ($\sim1\arcsec$)
object, brighter in H$\alpha$ and coincident with
the radio source. A spectrum obtained the same
night shows no continuum, but very strong nebular emission
lines. The object appears to be a very compact planetary nebula.
Its radio flux history is thus perplexing: 6.9~mJy at 6~cm in $\sim1990$,
rising to 10.5~mJy in 2005 and falling again to 5.8~mJy in 2006.
The 20~cm flux density in the MAGPIS survey (epoch 2001--2004)
is only 1.3~mJy, suggesting the source may be optically thick.
Further simultaneous multi-frequency observations are required to
measure the radio spectrum and derive clues as to the nature of the
source's variability.

\begin{figure*}
\epsscale{1.0}
\plotone{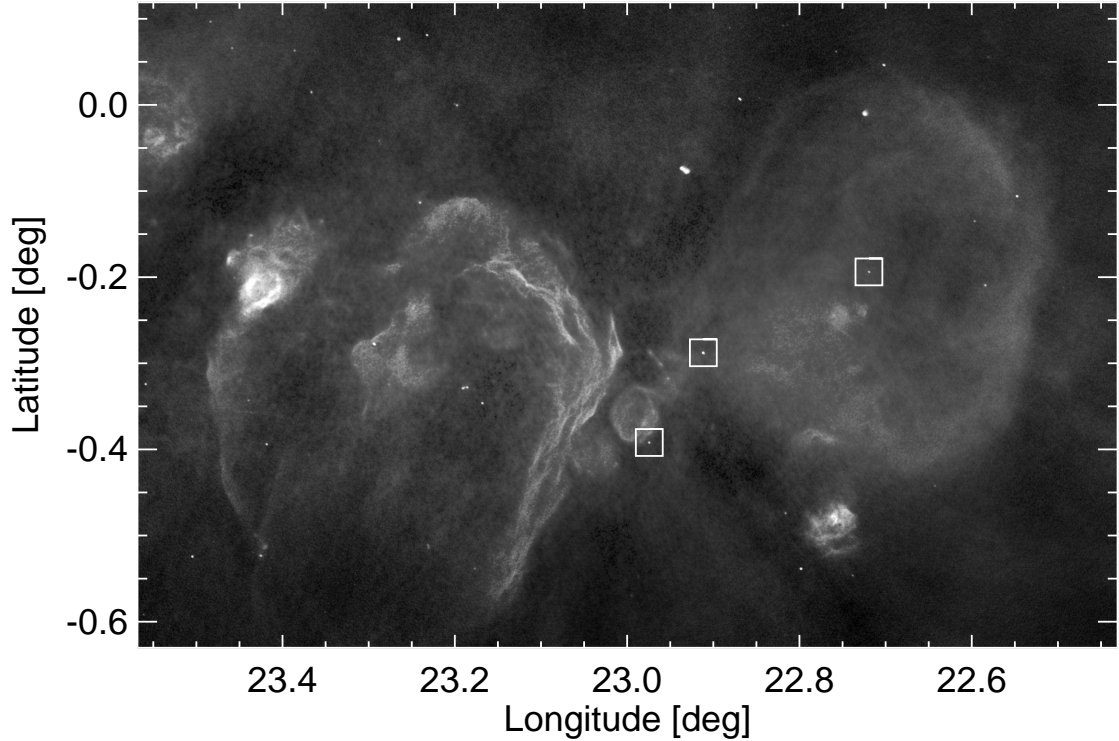}
\caption{
MAGPIS 20~cm image of supernova remnants W41 (G23.3${-}$0.3) and
G22.7${-}$0.2.  The boxes mark the positions of three
variable 6~cm sources (G22.7194${-}$0.1939, 
G22.9116${-}$0.2878, and G22.9743${-}$0.3920).
}
\label{fig-snrs}
\end{figure*}

\subsubsection{X-ray observations}

The brightest variable, G21.6552${-}$0.3611, is coincident with a
point-like X-ray source catalogued in the XMM Galactic Plane survey
(Hands et al.\ 2004). It has a hard band (2--6~keV) flux of 0.0051
ct s$^{-1}$ and is undetected in the soft (0.4--2.0~keV) band. For
an intrinsic power-law spectrum with spectral index $\Gamma = 1.9$,
the expected absorption column density through the Galactic plane
of $\sim10^{23}$ cm$^{-2}$ is consistent with the non-detection
in the soft band; the inferred intrinsic flux in the 0.2--10~keV
band would be $7 \times 10^{-13}$ erg cm$^{-2}$ s$^{-1}$. For an
extragalactic AGN at 1 Gpc, this corresponds to a luminosity of $8
\times 10^{42}$ erg s$^{-1}$, while for a Galactic object at 5~kpc,
the X-ray luminosity would be a modest $2 \times 10^{33}$ erg
s$^{-1}$; for a column density of only $10^{22}$ cm$^{-2}$, the
luminosity estimates are lower by a factor of 3. While this source
is the brightest of our variables, it has one of the lowest modulation
factors (decreasing by just $\sim50\%$ over 16 years). The
(non-contemporaneous) 20~cm flux density is lower than either of
the 6~cm values, suggesting a mildly inverted spectrum source. It
is not detected at any other wavelength. The most likely explanation
of this object is a flat-spectrum extragalactic radio source, one
of a handful we expect in our sample.

One other source, G30.4460${-}$0.2148, lies 27\arcsec\ from
the position of an ASCA Galactic Plane Survey catalog entry (Sugizaki
et al.\ 2001). The uncertainty in the X-ray position is 1\arcmin;
one other (brighter) radio source lies within the X-ray error circle
although at twice the distance from its centroid. The X-ray source
is a marginal detection ($4.6 \sigma$) with a 0.7--7.0~keV unabsorbed
flux of $2.6 \times 10^{-12}$ erg cm$^{-2}$ s$^{-1}$ for an intrinsic
power law index of $\Gamma = 1.9$ and an absorption column density
of $10^{23}$ cm$^{-2}$; the flux is roughly four times lower for
$N_H = 10^{22}$ cm$^{-2}$. Assuming the identification is correct,
the X-ray to radio flux ratio is thus $\sim20$ times greater than
our other X-ray detection, although still within the X-ray to radio
luminosity ratios characteristic of AGN. The primary distinguishing
feature, however, is that the radio source is coincident with a
very bright mid-IR source (saturated in all but the 3.6 $\mu$m band)
which is also detected at 1.1~mm.

The ASCA Galactic Plane Survey covered an area encompassing all but six of
our variables to a flux density level of approximately $10^{-12.5}$
erg cm$^{-2}$ s$^{-1}$; no other X-ray sources are coincident to
within 1\arcmin. The higher-resolution coverage of the Einstein,
ROSAT, XMM, and Chandra is much spottier; no further matches are
found within the 10\arcsec\ error circles of these other
catalogs.

\subsubsection{Low-frequency radio detections}

Three of the variable sources are detected at 90~cm. G22.9116${-}$0.2878
(Fig.~\ref{fig-snrs}) has a 90~cm flux density of $\sim180$~mJy; this is consistent with
a nonthermal spectral index of $\sim-0.9$ if one takes the most
recent (but far from contemporaneous) 6, 20, and 90~cm measurements.
The 6~cm flux density increased by more than a factor of three since
1990, making it one of the higher amplitude variables, but no other
information is available on this source. G30.6724${+}$0.9637 (the highest
latitude source detected) has a 90~cm flux density of $\sim70$
mJy, below that of the 20~cm flux density (90~mJy), possible
additional evidence for variability, as the 20:6~cm flux density
ratio is 3:1 (again, all non-contemporaneous). This is the smallest
amplitude variable in our sample and, given its distance from the
Galactic plane, an extragalactic counterpart is the most likely
explanation.

The third 90~cm detection is G22.7194${-}$0.1939, perhaps the most
intriguing source in our sample. An image of the region surrounding
this source is given in Figure~\ref{fig-snrs}. The source lies very near
to the geometric center of a 30\arcmin-diameter supernova remnant,
G22.7${-}$0.2 (Green 2004 and references therein) and 4\arcmin\ from
a fairly bright \ion{H}{2} region. There is no counterpart detected at mm,
IR, or optical wavelengths. The source brightened by a factor of
four between 2003 and 2006 at 6~cm; its 20~cm flux density is 12~mJy,
three times higher than the higher of the two 6~cm measurements.
The distance to the remnant is unknown, although its large angular
diameter would suggest it is not very remote (its diameter would
be $\sim45$~pc at 5~kpc). X-ray observations could reveal whether
or not this source is likely to be a compact object associated with
the supernova remnant.

\subsection{Summary}

We have discovered a relatively high surface density (2 deg$^{-2}$)
of variable radio sources in the Galactic plane and have argued
that the large majority of these ($\sim80$\%) are Galactic objects. While a few
are associated with young star formation activity, the identity
of the majority is unknown.
Follow up radio observations are required to confirm the variability
in these sources, establish the variability time scale(s), and
obtain contemporaneous spectral indices. Observations at optical,
infrared, and X-ray wavelengths could help establish counterparts
and identify the origin of the variable radio emission.

\acknowledgments

RHB and DJH acknowledge the support of the National Science Foundation
under grants AST-05-07598 and AST-02-6-55.  RHB's work was supported
in part under the auspices of the US Department of Energy by Lawrence
Livermore National Laboratory under contract W-7405-ENG-48.  DJH
was also supported in this work by NASA grant NAG5-13062.  RLW
acknowledges the support of the Space Telescope Science Institute,
which is operated by the Association of Universities for Research
in Astronomy under NASA contract NAS5-26555. The first three authors
are grateful for the hospitality of Quest University Canada
(\url{http://www.questu.ca}), an innovative new undergraduate university in
British Columbia, where this manuscript was completed.

\clearpage

\LongTables
\pagestyle{empty}
\advance\voffset by 7mm
\begin{landscape}
\begin{deluxetable}{ccccrrrcrrrcrrrccl}
\rotate
\tablenum{2}
\tablecolumns{18}
\tablewidth{0pc}
\tabletypesize{\scriptsize}
\tablecaption{Catalog of Variable 6~cm Galactic Plane Sources}
\tablehead{
& & &
\multicolumn{4}{c}{Epoch I (1990${+}$)\tablenotemark{a}} &
\multicolumn{4}{c}{Epoch II (2005)} &
\multicolumn{4}{c}{Epoch III (2006)} &
\colhead{Maximum\tablenotemark{b}} & \colhead{20 cm\tablenotemark{c}} & \\
\colhead{Name} & \colhead{RA} & \colhead{Dec} &
\colhead{Epoch} & \colhead{$S_p$} & \colhead{$S_i$} & \colhead{RMS} &
\colhead{Epoch} & \colhead{$S_p$} & \colhead{$S_i$} & \colhead{RMS} &
\colhead{Epoch} & \colhead{$S_p$} & \colhead{$S_i$} & \colhead{RMS} &
\colhead{Change} & \colhead{$S_p$} & \colhead{Comments\tablenotemark{d}} \\
& \colhead{(J2000)} & \colhead{(J2000)} &
& \colhead{(mJy)} & \colhead{(mJy)} & \colhead{(mJy)} &
& \colhead{(mJy)} & \colhead{(mJy)} & \colhead{(mJy)} &
& \colhead{(mJy)} & \colhead{(mJy)} & \colhead{(mJy)} &
\colhead{(rms)} & \colhead{(mJy)} & \\
\colhead{(1)} & \colhead{(2)} & \colhead{(3)} &
\colhead{(4)} & \colhead{(5)} & \colhead{(6)} & \colhead{(7)} &
\colhead{(8)} & \colhead{(9)} & \colhead{(10)} & \colhead{(11)} &
\colhead{(12)} & \colhead{(13)} & \colhead{(14)} & \colhead{(15)} &
\colhead{(16)} & \colhead{(17)} & \colhead{(18)} \\
}
\startdata
21.6552${-}$0.3611 & 18 31 57.508 & ${-}$10 11 22.43 & 2004.32 & 65.9 & 67.8 & 0.15 & \nodata & \nodata & \nodata & \nodata & 2006.53 & 43.0 & 43.2 & 0.31 & {\bfseries \em 66.6} & 22.7 & X-ray \\
22.7194${-}$0.1939 & 18 33 21.033 & ${-}$09 10 06.43 & 2003.59 & 1.1 & 0.8 & 0.14 & \nodata & \nodata & \nodata & \nodata & 2006.55 & 4.0 & 4.1 & 0.32 & \phn8.2 & 12.3 & 90 cm, SNR G22.7-0.2 \\
22.9116${-}$0.2878 & 18 34 02.837 & ${-}$09 02 28.03 & 1990.94 & 3.5 & 3.7 & 0.40 & \nodata & \nodata & \nodata & \nodata & 2006.55 & 11.4 & 16.5 & 0.32 & 15.4 & 46.2 & 90 cm \\
22.9743${-}$0.3920 & 18 34 32.343 & ${-}$09 02 00.69 & 1990.94 & 0.4 & 0.8 & 0.36 & \nodata & \nodata & \nodata & \nodata & 2006.55 & 7.2 & 5.4 & 0.33 & 13.3 & 14.7 &  \\
23.4186${+}$0.0090 & 18 33 55.619 & ${-}$08 27 15.92 & 1989.48 & 1.8 & 1.8 & 0.40 & \nodata & \nodata & \nodata & \nodata & 2006.55 & 4.6 & 5.6 & 0.32 & \phn5.4 & \phn2.0 &  \\
23.5585${-}$0.3241 & 18 35 23.014 & ${-}$08 29 01.42 & 1993.92 & $<$ 0.4 & \nodata & 0.21 & \nodata & \nodata & \nodata & \nodata & 2006.55 & 5.0 & 5.5 & 0.30 & 12.4 & 15.6 &  \\
23.6644${-}$0.0372 & 18 34 33.050 & ${-}$08 15 27.10 & 1989.89 & 5.3 & 5.4 & 0.18 & \nodata & \nodata & \nodata & \nodata & 2006.55 & 26.2 & 26.0 & 0.33 & 55.5 & \phn2.1 &  \\
24.3367${-}$0.1574 & 18 36 13.893 & ${-}$07 42 57.64 & 1997.03 & 2.0 & 2.2 & 0.25 & \nodata & \nodata & \nodata & \nodata & 2006.56 & 6.4 & 6.8 & 0.33 & 10.7 & 20.8 &  \\
24.5343${-}$0.1020 & 18 36 23.987 & ${-}$07 30 54.26 & 1990.50 & 0.8 & 2.8 & 0.52 & \nodata & \nodata & \nodata & \nodata & 2006.56 & 4.4 & 4.5 & 0.32 & \phn5.5 & \phn2.8 & GLIMPSE, MIPSGAL \\
24.5405${-}$0.1377 & 18 36 32.343 & ${-}$07 31 33.52 & 1990.86 & 0.7 & 1.5 & 0.36 & \nodata & \nodata & \nodata & \nodata & 2006.56 & 4.5 & 4.5 & 0.33 & \phn7.8 & \phn7.1 &  \\
25.2048${+}$0.1251 & 18 36 49.735 & ${-}$06 48 54.63 & 2001.41 & 6.6 & 6.6 & 0.22 & \nodata & \nodata & \nodata & \nodata & 2006.57 & 4.4 & 4.2 & 0.32 & {\bfseries \em \phn5.7} & \phn5.5 & GLIMPSE 8$\mu$m, MIPSGAL \\
25.4920${-}$0.3476 & 18 39 03.094 & ${-}$06 46 37.38 & 1990.93 & 4.3 & 4.3 & 0.17 & \nodata & \nodata & \nodata & \nodata & 2006.57 & 2.0 & 1.5 & 0.35 & {\bfseries \em \phn5.9} & \phn9.3 & embedded in nebula \\
25.7156${+}$0.0488 & 18 38 02.785 & ${-}$06 23 47.29 & 1991.62 & 2.1 & 2.6 & 0.38 & \nodata & \nodata & \nodata & \nodata & 2006.57 & 8.1 & 12.6 & 0.37 & 11.3 & 16.6 & GLIMPSE, MIPSGAL, Maser \\
26.0526${-}$0.2426 & 18 39 42.626 & ${-}$06 13 50.28 & 1991.61 & 7.5 & 7.4 & 0.33 & \nodata & \nodata & \nodata & \nodata & 2006.57 & 13.0 & 12.6 & 0.30 & 12.2 & 27.5 &  \\
26.2818${+}$0.2312 & 18 38 26.372 & ${-}$05 48 35.17 & 2000.85 & 9.5 & 9.0 & 0.27 & \nodata & \nodata & \nodata & \nodata & 2006.57 & 15.3 & 14.3 & 0.29 & 14.5 & 22.4 &  \\
27.8821${+}$0.1834 & 18 41 33.285 & ${-}$04 24 33.12 & 1991.89 & 3.8 & 3.3 & 0.21 & \nodata & \nodata & \nodata & \nodata & 2006.60 & 6.6 & 5.8 & 0.32 & \phn7.4 & \phn9.3 &  \\
28.6204${-}$0.3436 & 18 44 47.285 & ${-}$03 59 36.43 & 2002.42 & 2.8 & 3.3 & 0.18 & 2005.39 & 8.0 & 6.4 & 0.22 & 2006.60 & 8.6 & 8.4 & 0.31 & 18.2 & 35.4 &  \\
28.9841${-}$0.2947 & 18 45 16.772 & ${-}$03 38 51.21 & 1990.93 & 7.3 & 7.6 & 0.15 & 2005.24 & 5.8 & 5.8 & 0.23 & 2006.60 & 4.9 & 4.8 & 0.31 & {\bfseries \em \phn6.7} & 16.8 &  \\
29.0545${+}$0.8679 & 18 41 15.912 & ${-}$03 03 12.98 & \nodata & \nodata & \nodata & \nodata & 2005.20 & 8.3 & 8.1 & 0.24 & 2006.60 & 5.6 & 5.9 & 0.31 & \phn6.8 & \phantom{\tablenotemark{e}}26.9\tablenotemark{e} &  \\
29.1075${-}$0.1546 & 18 45 00.346 & ${-}$03 28 25.59 & 1991.49 & 6.6 & 6.5 & 0.13 & 2005.28 & 11.3 & 11.4 & 0.23 & 2006.60 & 10.9 & 10.5 & 0.33 & 17.9 & 14.0 &  \\
29.1978${-}$0.1268 & 18 45 04.317 & ${-}$03 22 50.49 & 2000.74 & 2.7 & 2.0 & 0.25 & 2005.27 & 4.4 & 4.1 & 0.22 & 2006.60 & 1.9 & 1.7 & 0.34 & \phn6.4 & \phn4.8 &  \\
29.2276${+}$0.5173 & 18 42 49.861 & ${-}$03 03 35.96 & \nodata & \nodata & \nodata & \nodata & 2005.20 & 19.2 & 20.1 & 0.22 & 2006.60 & 13.2 & 12.7 & 0.31 & 15.4 & 66.5 &  \\
29.4959${-}$0.3000 & 18 46 14.068 & ${-}$03 11 40.60 & 1990.93 & 5.3 & 5.2 & 0.16 & 2005.25 & 12.4 & 13.5 & 0.20 & 2006.60 & 6.9 & 7.5 & 0.32 & 27.4 & \phn5.0 &  \\
29.5779${-}$0.2685 & 18 46 16.334 & ${-}$03 06 26.06 & 1994.37 & 6.9 & 7.6 & 0.40 & 2005.22 & 10.5 & 11.5 & 0.24 & 2006.60 & 5.8 & 5.1 & 0.34 & 11.1 & \phn1.3 & PN, GLIMPSE, MIPSGAL, BGPS \\
29.6051${-}$0.8590 & 18 48 25.632 & ${-}$03 21 08.51 & \nodata & \nodata & \nodata & \nodata & 2005.20 & 1.8 & 2.5 & 0.23 & 2006.60 & 8.3 & 8.3 & 0.34 & 15.7 & \phantom{\tablenotemark{e}}44.2\tablenotemark{e} &  \\
29.7161${-}$0.3178 & 18 46 42.050 & ${-}$03 00 24.37 & 2002.14 & 15.9 & 16.6 & 0.26 & 2005.22 & 30.6 & 30.4 & 0.24 & 2006.60 & 28.7 & 30.1 & 0.37 & 41.2 & 25.6 &  \\
29.7195${-}$0.8788 & 18 48 42.423 & ${-}$03 15 34.64 & \nodata & \nodata & \nodata & \nodata & 2005.20 & 28.4 & 28.6 & 0.22 & 2006.60 & 20.3 & 19.5 & 0.32 & 20.7 & \phantom{\tablenotemark{e}}46.8\tablenotemark{e} &  \\
30.1038${+}$0.3984 & 18 44 51.460 & ${-}$02 20 05.69 & 2000.88 & 4.6 & 3.7 & 0.46 & 2005.20 & 6.3 & 6.8 & 0.20 & 2006.60 & 8.5 & 7.7 & 0.32 & \phn7.0 & \phn7.9 &  \\
30.4376${-}$0.2062 & 18 47 37.270 & ${-}$02 18 49.67 & 1990.94 & 0.9 & 1.2 & 0.27 & 2005.25 & 12.1 & 13.1 & 0.22 & 2006.60 & 16.8 & 17.5 & 0.32 & 38.1 & \phn8.7 &  \\
30.4460${-}$0.2148 & 18 47 40.045 & ${-}$02 18 37.04 & 1990.94 & 3.0 & 3.6 & 0.32 & 2005.24 & 5.9 & 6.6 & 0.22 & 2006.60 & 6.8 & 6.6 & 0.31 & \phn8.6 & \phn3.5 & GLIMPSE, MIPSGAL, BGPS, ASCA \\
30.6724${+}$0.9637 & 18 43 53.054 & ${-}$01 34 15.34 & \nodata & \nodata & \nodata & \nodata & 2005.20 & 28.0 & 29.1 & 0.32 & 2006.61 & 33.7 & 34.0 & 0.30 & 12.7 & \phantom{\tablenotemark{e}}89.9\tablenotemark{e} & 90 cm \\
31.1494${-}$0.1727 & 18 48 48.088 & ${-}$01 39 54.65 & 1995.00 & 1.8 & 1.6 & 0.26 & 2005.25 & 4.8 & 5.6 & 0.19 & 2006.61 & 3.2 & 1.8 & 0.35 & \phn9.0 & \phn3.1 &  \\
31.1595${+}$0.0449 & 18 48 02.703 & ${-}$01 33 24.86 & 2003.48 & 12.4 & 11.6 & 0.32 & 2005.33 & 16.7 & 18.8 & 0.23 & 2006.61 & 15.3 & 17.2 & 0.35 & 11.1 & 15.0 & GLIMPSE, MIPSGAL, BGPS, Maser \\
32.5898${-}$0.4468 & 18 52 24.293 & ${-}$00 30 29.63 & \nodata & \nodata & \nodata & \nodata & 2005.36 & 3.1 & 3.6 & 0.17 & 2006.62 & $<$ 0.7 & \nodata & 0.33 & \phn6.7 & \phn2.7 &  \\
32.7193${-}$0.6477 & 18 53 21.378 & ${-}$00 29 04.34 & \nodata & \nodata & \nodata & \nodata & 2005.35 & 3.6 & 4.3 & 0.17 & 2006.62 & 1.7 & 2.0 & 0.32 & \phn5.3 & \phn9.2 &  \\
37.2324${-}$0.0356 & 18 59 25.245 & ${+}$03 48 37.48 & 2000.67 & 2.8 & 2.4 & 0.27 & \nodata & \nodata & \nodata & \nodata & 2006.63 & $<$ 0.5 & \nodata & 0.27 & {\bfseries \em \phn5.8} & 14.7 &  \\
37.7347${-}$0.1126 & 19 00 36.987 & ${+}$04 13 18.60 & 2003.78 & $<$ 1.3 & \nodata & 0.63 & \nodata & \nodata & \nodata & \nodata & 2006.63 & 11.3 & 12.2 & 0.32 & 14.2 & 11.0 & MIPSGAL, BGPS, Maser \\
37.7596${-}$0.1001 & 19 00 37.037 & ${+}$04 14 59.08 & 2002.39 & $<$ 1.5 & \nodata & 0.74 & \nodata & \nodata & \nodata & \nodata & 2006.63 & 11.5 & 12.0 & 0.32 & 12.4 & 18.5 &  \\
39.1105${-}$0.0160 & 19 02 47.984 & ${+}$05 29 21.52 & 2001.14 & 0.8 & 0.9 & 0.19 & \nodata & \nodata & \nodata & \nodata & 2006.65 & 3.1 & 4.2 & 0.32 & \phn6.3 & 12.9 &  \\
\enddata
\tablenotetext{a}{The mean epoch of observation for the Epoch I data (as given by the ``Epoch'' column)
    span a wide range; about 70\% of the observations were taken in 1989 and 1990, and the remaining
    30\% in 2004.  Mean epochs between those values indicate that observations from both epochs contribute to the
    coadded image.}
\tablenotetext{b}{Greatest brightness difference between two epochs normalized by the rms flux errors.  Bold
text indicates objects brightest at epoch 1 (possibly contaminated by resolution effects; see discussion for
details).}
\tablenotetext{c}{1.4 GHz flux density from MAGPIS catalog (Helfand et al.\ 2006) unless otherwise noted.}
\tablenotetext{d}{Notes on detections at other wavelengths:
    90~cm (Helfand et al.\ 2006; Brogan et al.\ 2005),
    GLIMPSE (3.6, 4.5, 5.8 and 8.0 $\mu$m unless wavelength is noted; Benjamin et al.\ 2003),
    MIPSGAL (24 $\mu$m; Carey et al.\ 2009),
    BGPS (Bolocam GPS 1.1~mm; Aguirre et al.\ 2009),
    Maser (ultracompact H~II regions with 6.7 GHz methanol maser; Pestalozzi et al.\ 2005, Pandian et al.\ 2007),
    X-ray ({\it XMM-Newton} Galactic Plane Survey, Hand et al.\ 2004),
    ASCA (X-ray; Sugizaki et al.\ 2001), PN (compact planetary nebula; this paper).
    }
\tablenotetext{e}{Source is outside MAGPIS survey area; 1.4 GHz flux from White et al.\ (2005).}
\end{deluxetable}

\clearpage
\end{landscape}

\end{document}